\documentclass[12pt]{article}

\usepackage{graphicx}
\usepackage{amsmath}
\usepackage{amssymb}
\allowdisplaybreaks

\begin{document}

\baselineskip=17.5pt plus 0.2pt minus 0.1pt

\renewcommand{\theequation}{\arabic{equation}}
\renewcommand{\thefootnote}{\fnsymbol{footnote}}
\makeatletter
\def\CR{\nonumber \\}
\def\pt{\partial}
\def\be{\begin{equation}}
\def\ee{\end{equation}}
\def\bea{\begin{eqnarray}}
\def\eea{\end{eqnarray}}
\def\eq#1{(\ref{#1})}
\def\la{\langle}
\def\ra{\rangle}
\def\hyp{\hbox{-}}


\begin{titlepage}
\title{\hfill\parbox{4cm}{ \normalsize YITP-07-61}\\
\vspace{1cm} The lowest modes around Gaussian solutions \\ 
of tensor models and the general relativity}
\author{
Naoki {\sc Sasakura}\thanks{\tt sasakura@yukawa.kyoto-u.ac.jp}
\\[15pt]
{\it Yukawa Institute for Theoretical Physics, Kyoto University,}\\
{\it Kyoto 606-8502, Japan}}
\date{}
\maketitle
\thispagestyle{empty}
\begin{abstract}
\normalsize
In the previous paper, the number distribution of the low-lying spectra 
around Gaussian solutions representing various dimensional fuzzy tori
of a tensor model was numerically shown to be in accordance with
the general relativity on tori.
In this paper, I perform more detailed numerical analysis of the properties
of the modes for two-dimensional fuzzy tori,
and obtain conclusive evidences for the agreement.
Under a proposed correspondence between the rank-three tensor in tensor models 
and the metric tensor in the general relativity,
conclusive agreement is obtained between
the profiles of the low-lying modes in a tensor model 
and the metric modes transverse to the general coordinate transformation.
Moreover,
the low-lying modes are shown to be 
well on a massless trajectory with quartic momentum dependence in the tensor model. 
This is in agreement with that
the lowest momentum dependence of metric fluctuations in the general relativity
will come from the $R^2$-term, since the $R$-term is topological in two dimensions.
These evidences support the idea that the low-lying low-momentum dynamics around the 
Gaussian solutions of tensor models is described by the general relativity.
I also propose a renormalization procedure for tensor models.
A classical application of the procedure makes the patterns of the 
low-lying spectra drastically clearer, 
and suggests also the existence of massive trajectories.
\end{abstract}
\end{titlepage}

\section{Introduction}
\label{sec:intro}
Thought experiments in combination of quantum mechanics and general relativity \cite{Garay}-\cite{Maziashvili:2006ey}
and also in string theory \cite{Yoneya:2000bt} show various bounds on accuracy of 
space-time measurements. 
The existence of such bounds implies that
space-time can never be observed as a smooth continuous manifold,
which is the classical space-time notion in general relativity.
Moreover the obvious distinction between space-time and matter fields on it 
would be an obstacle in pursuit of unification.
Especially, it would be nice if space-time and metric tensor 
can merely be regarded as different sides of a single object.
Thus search for a satisfactory alternative to 
the classical space-time notion has theoretical interests.

The notion of fuzzy space would be a possible candidate for such an 
alternative \cite{snyder}-\cite{Ho:2007vk}. 
A fuzzy space is defined by the algebra of functions on it, which is allowed to
be not only noncommutative but also nonassociative.
Therefore a fuzzy space is a generalized notion 
which includes both the noncommutative 
\cite{snyder}-\cite{Balachandran:2005ew} and the nonassociative 
\cite{Jackiw:1984rd}-\cite{Ho:2007vk} spaces. 
Since an algebra can be defined by a rank-three 
tensor\footnote{As in Equation~\eq{eq:defofc}.}, 
dynamical models of a rank-three tensor, which are denoted 
by tensor models \cite{Ambjorn:1990ge}-\cite{Freidel:2005cg} in this paper, 
may be regarded as dynamical theory of fuzzy spaces \cite{Sasakura:2005js}.
An important advantage of the formalism is that fuzzy analogue of 
the general coordinate transformation can easily be embedded into such tensor models. 
Another one is that one can treat various types of spaces in a unified manner,
e.g. irrespective of topologies and dimensions.  
This possibility to describe dynamical fuzzy spaces in terms of tensor models
has been pursued by the present author, and some successful results 
have been obtained. In \cite{Sasakura:2005js,Sasakura:2005gv,Sasakura:2006pq,Sasakura:2007sv}, various classical solutions of tensor models
representing fuzzy spaces with some physical interests
have been obtained, and their properties have been analyzed. Especially in the previous paper \cite{Sasakura:2007sv}, 
the number counting of the low-lying fluctuation modes around a specific class of
solutions, which represent various dimensional fuzzy flat tori and 
are denoted by Gaussian solutions in this paper, agrees with
what is expected from the general relativity.
This result suggests that the effective low-lying physics 
around the Gaussian solutions can be described by the general relativity. 
If this truly holds, metric tensor and space come from a single object
as mentioned in the preceding paragraph, because the rank-three tensor 
is the only dynamical variable of tensor models.

Although the agreement is non-trivially realized 
in various dimensions, it is obvious that 
such number counting is not enough for a definite conclusion.
The main purpose of the present paper is to go beyond the number counting 
to conclusively show the agreement.    
The paper is organized as follows. 
In Section~\ref{sec:tensormodels}, fuzzy space, tensor models, and 
Gaussian solutions are reviewed. A tensor model with Gaussian solutions,
which will be analyzed throughout the present paper, is given.
In Section~\ref{sec:cor}, a proposal is given 
of a correspondence between the rank-three tensor of tensor models 
and the metric tensor in the general relativity.
Then the DeWitt supermetric with its unique parameter being fixed 
is derived from a quadratic invariant metric of tensor models. 
The supermetric determines the metric fluctuations transverse to 
the general coordinate transformation.
In Section~\ref{sec:numerical}, numerical analysis of the fluctuation 
modes is performed for two-dimensional fuzzy flat tori.
Conclusive agreement is obtained between the transverse metric fluctuations
and the low-lying fluctuation modes of the tensor model. 
The low-lying spectra are observed to form a massless trajectory with quartic
momentum dependence.
In Section~\ref{sec:ren}, a renormalization procedure for tensor models is 
proposed. 
A classical application is shown to make much clearer the patterns of 
the low-lying spectra, and also shows the existence of a massive trajectory well over the massless one.
The final section is devoted to summary, conclusions, and discussions.

\section{Tensor models}
\label{sec:tensormodels}
\subsection{Fuzzy spaces and tensor models}
The idea of regarding tensor models as dynamical theory of fuzzy spaces 
was originally presented in \cite{Sasakura:2005js}.
In this subsection, I will recapitulate the discussions, stressing more on
physical motivations.

Before describing a fuzzy space, 
let me start with a usual space with coordinates,  
$x^\mu\ (\mu=1,\cdots,D)$. Let me only consider the {\it real}\, functions
on the space. One will be able to take a basis of the space of 
all the real functions as
$\{f_a(x)\}$ 
with an appropriate index set. In some cases, one can choose a 
basis with indices having some physical meanings such as frequencies, but 
in general the indices are merely abstract labels.  
The algebraic relations among the functions can be 
parameterized by a real rank-three tensor ${C_{ab}}^c$ as
\be
\label{eq:defofc}
f_a(x)\, f_b(x)={C_{ab}}^c\, f_c(x).
\ee
Introducing a rank-two tensor $g_{ab}$ defined by
\be
\label{eq:gusual}
g_{ab}=\int d^Dx\, f_a(x) f_b(x),
\ee
one obtains
\be
\label{eq:cusual}
C_{abc}=\int d^Dx \, f_a(x)f_b(x)f_c(x),
\ee
where $C_{abc}={C_{ab}}^{c'}g_{c'c}$. 

Here both $g_{ab}$ and $C_{abc}$ are real symmetric tensors, 
because of the associativity and commutativity of the products
among the functions on a usual space.
It is important to note that one can freely take another basis for the 
function space. Since such a change of basis can be described by 
a general linear transformation, tensors related by a general linear transformation
describe an identical space.
The general coordinate transformation can be embedded into the general
linear transformation, because any function can be expressed as  
a linear combination of the basis functions. Namely, there always exist  
$M_a{}^b$ such that
$f_a(x')={M_a}^bf_b(x)$, where $x'=x'(x)$ is a transformed coordinate 
and ${M_a}^b$ are the numerical coefficients of expansion.

If $g_{ab}$ and $C_{abc}$ are given and $g_{ab}$ is invertible, one can get
${C_{ab}}^c$ in the product rule \eq{eq:defofc}. 
As will be explained later at the end of this subsection,
it is more convenient to consider  
the two tensors, $g^{ab}=(g^{-1})^{ab}$ and $C_{abc}$, as 
the fundamental variables 
of tensor models rather than ${C_{ab}}^c$ itself in the product rule.
The main reason comes from the requirement of 
the invariance of an action under the general linear transformation, as 
will be explained at the end of this subsection.

In field theory, the basic variables are fields, which are 
similar to functions. Their product and integration over a space-time
are basic operations necessary in extracting physics from field theory, so that
the tensors $g^{ab}$ and $C_{abc}$ 
are more directly related to observables in field theory
than a coordinate system itself. Therefore it would be acceptable
in physics to describe a space in terms of these tensors, 
but not by coordinates.

A fuzzy space may be defined by a deformation of the above tensors away from
the values corresponding to a usual space. 
The type of deformation relevant in this paper is basically a kind of truncation 
of the function space by introducing a cutoff to high frequency modes without
changing the symmetric properties of the tensors $g^{ab}$ and $C_{abc}$ under 
exchange of the indices.
The cutoff may also be introduced by a cutoff function which  
vanishes gradually at high frequencies.
Such truncation of high frequency modes will make it hard to distinguish 
nearby points, which makes a space to become ``fuzzy".  

Because the symmetry under exchange of the indices of the tensors is assumed 
in the deformation, 
the product rule \eq{eq:defofc} remains
commutative. However, this kind of deformation generally introduces 
nonassociativity into the algebra \eq{eq:defofc}, 
because truncation of modes breaks closure of an original algebra.
The physical importance of this kind of nonassociative fuzzy spaces
comes from the fact that metric can be incorporated into the function algebra, 
as will explicitly be presented in Section~\ref{sec:cor}.
This is in sharp contrast with a noncommutative space 
\cite{snyder}-\cite{Balachandran:2005ew}, where
metric is given by an additional element, Laplacian \cite{connes}. 

The above discussions naturally lead one to suspect that a dynamical theory of 
fuzzy spaces may be constructed as a dynamical theory of the 
symmetric tensors $g^{ab}$ and $C_{abc}$, and may 
reproduce the general relativity in a certain limit. Since 
a fuzzy space is invariant under the transformation of a basis, 
tensor models must have a symmetry under the general linear transformation,
\bea
C_{abc}&=&{M_a}^{a'}{M_b}^{b'}{M_c}^{c'}C'_{a'b'c'}, \cr
g^{ab} &=& {(M^{-1})_{a'}}^a {(M^{-1})_{b'}}^b g'{}^{a'b'},
\eea 
which is a fuzzy analogue of the general coordinate transformation on a usual space.
Therefore an action of a tensor model is a function of $g^{ab}$ and $C_{cde}$,
\be
\label{eq:generalaction}
S(g^{ab},C_{cde}),
\ee 
where the upper and lower indices are contracted for the symmetry requirement.
In principle, 
it is possible to take another set of dynamical variables than $g^{ab}$ and $C_{abc}$.
However the above choice seems to be the simplest. 
For example, one would try to take ${C_{ab}}^c$ as its only dynamical variable. 
Then to construct an invariant, one would need a tensor with more upper indices
than lower ones. One would be able to 
make such a tensor by an inverse of a matrix ${C_{ab}}^c{C_{a'c}}^b$,
but this would lead to an action singular at the non-invertible values.

\subsection{Euclidean models}
The general discussions on tensor models in the previous subsection 
seem to favor the existence of two kinds of symmetric tensors $g^{ab}$ and $C_{abc}$ 
as dynamical variables in tensor models. 
However, in the actual numerical analysis,
treating both as dynamical variables would become too complicated.
Also from theoretical viewpoints, 
the existence of two independent tensors would introduce
more ambiguities into the formalism.
Therefore a Euclidean tensor model \cite{Sasakura:2005gv,Sasakura:2007sv}, 
which has the non-dynamical rank-two tensor fixed at $g^{ab}=\delta^{ab}$, 
will be considered throughout this paper.

It is important to note that such a model is only invariant under an
orthogonal subgroup of the general linear transformation.
Therefore the direct link between the symmetry of a tensor model
and the general coordinate transformation is missing in such a model.   
 However there seem to exist several reasons for the possibility that one may 
ignore $g^{ab}$ as dynamical variables without changing the essential features
of the system as listed in the following.
Especially
it is noteworthy that, because of the following first reason, 
Euclidean tensor models are, so called, background independent theory of 
space.
\begin{itemize}
\item
Under the assumption that $g^{ab}$ is positive definite as a matrix, 
it can be diagonalized to 
$g^{ab}=\delta^{ab}$ by the general linear transformation. This is a partial gauge 
fixing of the general linear transformation to the orthogonal subgroup.
Therefore, classically, there are no distinctions between a full tensor model and 
a Euclidean one under the positivity 
assumption\footnote{Quantum mechanically, however, one would need to take into 
account the contributions from the FP ghosts associated with
the partial gauge fixing.}.  
In fact, the correspondence to the general relativity \eq{eq:correspondence},
which will be discussed later, does not require $g^{ab}$ to be varied. 
\item
In the previous paper \cite{Sasakura:2007sv}, it is numerically shown 
that the distribution of the low-lying modes around the Gaussian classical solutions 
of a Euclidean tensor model agrees with the general relativity. 
\item
What is really required in a tensor model would be a symmetry which resembles
the general coordinate transformation at low-momentum modes. In general, 
there are no clear distinctions 
between the general linear and the orthogonal transformations, 
if the two are compared only at a small window of low-momentum modes. 
\item
The number of components of $g^{ab}$ is negligible in comparison 
with $C_{abc}$ in the limit $n\rightarrow \infty$,
where $n$ is the total number of the possible values of an index, 
e.g. $a=1,2,\cdots,n$. 
Therefore $g^{ab}$ would not play essential roles in the dynamics in the limit.
\end{itemize}
Thus, in the following discussions, only a Euclidean tensor model,
\be
\label{eq:etensaction}
S(g^{ab}=\delta_{ab},C_{abc}),
\ee 
will be considered. 

\subsection{A Euclidean tensor model with Gaussian classical solutions}
\label{sec:themodel}
The orthogonal group symmetry of Euclidean models is so loose that
the explicit form of an invariant action \eq{eq:etensaction} has infinite
possibilities. In general the dynamics of the models will heavily depend on its
specific form, and the models have no powers to predict any quantum gravitational
phenomena.
Presently no principles are known to constrain the choices,
but better understanding of the properties of the models might finally 
lead to a hint for a preferred formulation. 
It might also be possible that the models can be classified into a finite number of 
universality classes, when the thermodynamics or quantum properties are 
investigated.

In this regard, 
an obvious direction of study is to consider the simplest choices of the actions. 
In the papers \cite{Sasakura:2005js,Sasakura:2005gv,Sasakura:2006pq}, some
actions with quadratic and quartic terms in $C_{abc}$ are considered.
Especially in \cite{Sasakura:2006pq}, 
considered is an action the equation of motion of which has  
various commutative but nonassociative deformations of usual spaces, 
tori and spheres of various dimensions, as classical solutions.
It is interesting that a single equation of motion contains various physically
meaningful solutions corresponding to spaces 
with various topologies, dimensions, curvatures and sizes.

Another direction which has been pursued is to look for relations between
tensor models and the general relativity.
In the previous paper \cite{Sasakura:2007sv}, 
the fluctuation spectra around a specific type of 
solutions, the Gaussian solutions, to a tensor model for 
one- to four-dimensional tori are numerically studied.
It was shown that 
the number of the low-lying low-momentum modes at each momentum sector 
agrees exactly with what is expected from the general relativity.

It is obvious that such  number counting is not enough to definitely 
prove the relations, and more detailed comparisons are required.
In fact, as will be discussed in Section~\ref{sec:correspondence}, 
the Gaussian solutions have a natural generalization to 
incorporate a correspondence between the rank-three tensor in tensor models 
and the metric tensor in the general relativity. Moreover,
they are right on the trajectory of the renormalization procedure proposed in
Section~\ref{sec:renproc}, and can be expected to play significant roles 
in the dynamics.
Therefore the Gaussian solutions seem to be interesting backgrounds to 
perform more detailed comparisons with the general relativity. 
However, the action having such Gaussian solutions considered in the previous 
paper \cite{Sasakura:2007sv} is very complicated, and 
it is hard to obtain fully reliable numerical results from the model
because of the  heavy requirement of machine powers. Therefore,
in the following, a much more simpler model having the Gaussian solutions will be
given. This model contains a fractional inverse power of a matrix, 
and therefore cannot be considered as a well-defined action for all the 
values of $C_{abc}$. 
But, as for small fluctuations around the Gaussian solutions, 
this irregularity will not make any harms as can be seen later.

The Gaussian solutions have the following Gaussian form,
\bea
\label{eq:cx}
\bar 
C_{x_1,x_2,x_3}&=&B \exp\left[ -\beta \left( (x_1-x_2)^2+(x_2-x_3)^2+(x_3-x_1)^2\right)\right], \cr
g^{x_1,x_2}&=& \delta^D(x_1-x_2),
\eea
where $B,\beta$ are positive numerical coefficients, 
$x_i$ are $D$-dimensional continuous coordinates,
$x_i=(x_i^1,x_i^2,\cdots,x_i^D)$, and  $(x)^2$ is a short-hand notation 
for $\sum_{\mu=1}^D (x^\mu)^2$. Only the Euclidean signature of space 
is considered in this paper. 
The integration measure for the contraction of indices is defined by
$\int d^Dx$. 
As in \eq{eq:cx}, throughout this paper,
the symbol $\bar{}$ on a tensor is used to represent the solution, which is 
distinguished from the dynamical variable of a tensor model. 
As can be checked easily,
the algebra of functions defined by \eq{eq:cx}, 
$f_{x_1}f_{x_2}={ \bar C_{x_1,x_2}}{}^{x_3} f_{x_3}$, 
is commutative nonassociative, and approaches 
$f_{x_1}f_{x_2}\propto \delta^D (x_1-x_2) f_{x_1}$ in the limit $\beta\rightarrow \infty$.
Noting that \eq{eq:cx} has the Poincare symmetry, and that 
the products of functions defined by $f_{x_i}\equiv\delta^D(x-x_i)$ 
satisfies $f_{x_1}f_{x_2}= \delta^D (x_1-x_2) f_{x_1}$ on a usual continuous space,
the fuzzy space defined by \eq{eq:cx} for a finite $\beta$ 
can be regarded as 
a commutative nonassociative deformation of a usual $D$-dimensional flat space. 
This fuzzy space is essentially the same one considered in \cite{Sasai:2006ua}
to investigate the one-loop properties of field theory on nonassociative space.
It is important to note that, because of the identity form of $g^{x_1,x_2}$,
\eq{eq:cx} has a gauge fixed form which 
can be embedded into a Euclidean tensor model.

In the actual computation of contracting indices, however, 
the representation of \eq{eq:cx} with  the indices of 
the spatial coordinates $x$ 
is inconvenient, because the momentum conservation coming from the 
translational symmetry of the solution
is not obvious.
One may obtain an expression in terms of momentum indices
by insertion of the identity,
\be
\delta^D(x-y)=\frac{1}{(2\pi)^D}\int d^Dp \exp\left(ip(x-y)\right), 
\ee
to all the contractions of the index $x$. This is equivalent to the Fourier
transformation ${F_p}^x=\frac1{(2\pi)^{D/2}}\exp(i p x)$ for a lower index of $x$, 
and its inverse transformation ${F_x}^p=\frac1{(2\pi)^{D/2}}\exp(-i p x)$
for an upper index. Performing this index transformation, one obtains the Gaussian 
solutions in the momentum representation as  
\bea
\label{eq:cp}
\bar C_{p_1,p_2,p_3}&=&A \,
\delta^D(p_1+p_2+p_3) \exp \left[ -\alpha \left( p_1^2+p_2^2+p_3^2\right) \right],
\cr
g^{p_1,p_2}&=&\delta^D(p_1+p_2),
\eea
where $A$ and $\alpha\sim \frac{1}{\beta}$ are positive numerical constants. 
The integration measure for the contraction is defined by $\int d^Dp$.
Since the transformation is not real valued, it is not to take another gauge
in a tensor model, but is rather to transform to 
a technically convenient representation for actual computations. 
Therefore, on every occasion, one has to take care of how the reality condition 
is transformed. On $C_{p_1,p_2,p_3}$, this is
\be
\label{eq:cstar}
C_{p_1,p_2,p_3}=C^*_{-p_1,-p_2,-p_3},
\ee
where $*$ denotes the complex conjugation.

To construct a tensor model with the Gaussian solutions \eq{eq:cp},
I follow the same procedure as was performed 
in the previous paper \cite{Sasakura:2007sv}, i.e.
constructing first an equation of motion with the Gaussian solutions, and 
defining an action by its square. The reason behind for this adhoc easy way to 
define an action comes from the idea that the general relativity can be regarded as 
low-energy effective phenomena associated with the spontaneous breakdown of 
the local translational symmetry \cite{Borisov:1974bn}. 
In their discussions, the derivation of the general relativity is based on
the analysis of the non-linear realization of the local coordinate transformation
rather than starting from an action. 
This symmetry breaking is very similar to what occurs in tensor models, that is,
the $O(n)$ symmetry, the fuzzy analogue of the general coordinate transformation,
breaks down at classical solutions \cite{Sasakura:2007sv}. 
Therefore one may guess that the essential properties of the 
low-lying spectra are determined by the symmetry breakdown, but not by
the details of an action, although the precise correspondence to the 
continuum discussions is missing presently.

Let me compute the following two tensors,
\bea
\label{eq:bark}
{\bar K_{p_1}}{}^{p_2}&\equiv & \bar C_{p_1,p_3,p_4}\bar C^{p_2,p_3,p_4}=A^2
\left( \frac{\pi}{4 \alpha}\right)^{\frac{D}{2}}\exp\left(-3 \alpha p_1^2\right)
\delta^D(p_1-p_2), \\
\label{eq:barh}
\bar H_{p_1,p_2,p_3}&\equiv&{\bar C_{p1,p4}}{}^{p_5}{\bar C_{p2,p5}}{}^{p_6} 
{\bar C_{p_3,p_6}}{}^{p_4}\cr 
&=&A^3 \left(\frac{\pi}{6\alpha}\right)^{\frac{D}{2}}
\exp\left(-\frac{5\alpha}{3}\left(p_1^2+p_2^2+p_3^2\right)\right)
\delta^D(p_1+p_2+p_3).
\eea
Since $\bar K_{p_1}{}^{p_2}$ has a diagonal form, one can safely consider its fractional power\footnote{The $\delta^D(p_1-p_2)$ is the identity matrix, any fractional power of which remains the same.},
\be
(\bar K^{-\frac29})_{p_1}{}^{p_2}=
A^{-\frac49}\left( \frac{\pi}{4 \alpha}\right)^{-\frac{D}{9}}
\exp\left(\frac23\, \alpha\, p_1^2\right) \delta^D(p_1-p_2).
\ee 
Therefore the tensor $W_{p_1,p_2,p_3}$ defined by
\be
W_{p_1,p_2,p_3}=C_{p_1,p_2,p_3}- 
(K^{-\frac29})_{p_1}{}^{p_1'}(K^{-\frac29})_{p_2}{}^{p_2'}(K^{-\frac29})_{p_3}{}^{p_3'}
H_{p_1',p_2',p_3'}
\ee
vanishes at \eq{eq:cp}, if 
\be
A=\left(\frac{27 \alpha}{2\pi}\right)^\frac{D}{4},
\ee
where $K,H$ are defined by omitting $\bar{\ }$ in \eq{eq:bark} and \eq{eq:barh},
\bea
\label{eq:defk}
{ K_{a}}{}^{b}&\equiv &  C_{acd}C^{bcd}, \\
\label{eq:defh}
 H_{abc}&\equiv&{ C_{ad}}{}^{e}{ C_{be}}{}^{f} 
{C_{cf}}{}^{d},
\eea
respectively.
Thus I consider 
\be
\label{eq:generaleq}
W_{abc}=0
\ee
as the equation of motion in this paper. 
It should be noted that the equation of motion contains the Gaussian solutions
irrespective of the dimension $D$, and is therefore giving a kind of unified
description to all the Gaussian solutions.

The simplest action which has \eq{eq:generaleq} as its equation of motion can be
given by 
\be
\label{eq:theaction}
S=W_{abc}W^{abc}.
\ee
It should be noted that the action \eq{eq:theaction} is semi-positive definite, and 
if a solution to \eq{eq:generaleq} is found, this is necessarily a stable solution.
Therefore there will exist no negative spectra of quadratic fluctuations around it.

\section{Correspondence between the rank-three tensor and the metric}
\label{sec:cor}
In the previous paper \cite{Sasakura:2007sv}, it was numerically shown that 
the number of the low-lying spectra at each momentum sector
around the Gaussian solutions representing 
one- to four-dimensional fuzzy tori  agrees with the 
expectation from the general relativity.
For two-dimensional tori, it was observed that 
the number of low-lying spectra is three at $p=(0,0)$, and 
one at the other sectors. 
Below I will repeat the explanation of the number distribution in
the general relativity.

The actual degrees of freedom of metric must be evaluated 
modulo the general coordinate transformation.
The infinitesimal local coordinate transformation of the metric
tensor\footnote{This should not be confused with $g^{ab}$ in 
tensor models.}  $g_{\mu\nu}(x)$
is given by $\delta g_{\mu\nu}=\nabla_\mu v_\nu+\nabla_\nu v_\mu$,
where $v_\mu(x)$ is a local translation vector. On a flat torus, the transformation
is expressed as 
\be
\label{eq:ggauge}
\delta g_{\mu\nu}(p)=i p_\mu v_\nu(p)+i p_\nu v_\mu(p)
\ee
in the momentum representation. 
Since the gauge transformation is null at $p=(0,0)$ sector,
there remain all the components of the metric tensor as the 
actual degrees of freedom, the number of which is given by $D(D+1)/2$.
On the other hand, at $p\neq(0,0)$ sectors, 
this must be subtracted by the number of components of $v_\mu$, and  
the number of the actual degrees of freedom becomes $D(D-1)/2$.
For $D=2$, these numbers are 3 and 1, respectively, and agree with the numerical 
results mentioned above. 

It is obvious that such comparison of the numbers is not enough to definitely
prove the relation between tensor models and general relativity.
Therefore the main purpose of this paper is to compare the details of the corresponding
modes in tensor models and general relativity.
In the followings, I will first propose a correspondence between the rank-three tensor
of tensor models and the metric tensor in general relativity.
Then, in a slow varying approximation, an $O(n)$-invariant quadratic measure 
of tensor models will lead to the DeWitt supermetric \cite{DeWitt:1962ud} with its unique parameter being fixed.
This supermetric will determine the metric fluctuations transverse to
the local coordinate transformation. 
Then these transverse metric fluctuations will determine 
the profiles of the fluctuation modes of tensor models under the correspondence, 
which should be compared with the numerical analysis.

\subsection{The correspondence}
\label{sec:correspondence}
A proposal of correspondence 
between the rank-three tensor and the metric tensor can be 
obtained by an invariant generalization of the solution \eq{eq:cx},
\bea
\label{eq:correspondence}
C_{x_1,x_2,x_3}&=&B g(x_1)^\frac14 g(x_2)^\frac14 g(x_3)^\frac14
\exp\left[ - \beta \left( d(x_1,x_2)^2+d(x_2,x_3)^2+d(x_3,x_1)^2\right)\right], \cr
g^{x_1,x_2}&=& \delta^D(x_1-x_2),
\eea
where $g(x)={\rm det}\left(g_{\mu\nu}(x)\right)$ and $d(x,y)$ denotes
the distance\footnote{The distance between a pair of points may be defined by 
the length of the shortest path between them.
The definition may become singular for topological reasons or large fluctuations
of $g_{\mu\nu}$, but will cause no problems if only small fluctuations from
a background are considered as in this paper.} between the two points $x,y$,
where the infinitesimal length is defined by $ds^2=g_{\mu\nu}dx^\mu dx^\nu$ as usual.
The integration measure in contraction of indices is defined by $\int d^Dx$.
The parameter $\beta$ is redundant in the sense that it can be absorbed into 
$g_{\mu\nu}$.  But here it is left for the later convenience to represent the order of
$g_{\mu\nu}$, namely $O(g_{\mu\nu})\sim 1$ is kept.  
The correspondence \eq{eq:correspondence} 
is only applicable in the vicinity of the solution \eq{eq:cx},
but will be enough to analyze the small fluctuations around it. 
The expression \eq{eq:correspondence} respects the invariance under the general 
coordinate transformation, that is an invariant tensor such as $C_{abc}C^{abc}$ is
invariant under the general coordinate transformation. 
This is because the distance $d(x,y)$ is invariant from its definition, 
and the integration over $x$, which appears in the contraction of an index,
\be
C_{a,b,x}C^{x,c,d}=\int d^Dx \sqrt{g(x)}\cdots,
\ee
is also invariant.
It is very interesting to see that,
since $g^{x_1,x_2}$ in \eq{eq:correspondence} does not depend on the metric tensor,
the general coordinate transformation is realized 
within Euclidean models. 

\subsection{An $O(n)$-invariant measure and the DeWitt supermetric} 
The measure of tensor models which was used in the numerical analysis 
in the previous paper \cite{Sasakura:2007sv} and will be used throughout this paper 
is the quadratic $O(n)$-invariant measure\footnote{
The measure may generally be added by $\delta C_{ab}{}^b\,\delta C_c{}^{ac}$, 
which will shift the
parameter of the DeWitt supermetric discussed below.},
\be
\label{eq:Cmeasure}
ds_C^2=\delta C_{abc}\, \delta C^{abc}.
\ee
The corresponding measure in the space of $g_{\mu\nu}(x)$ can be obtained by
putting the correspondence \eq{eq:correspondence} into \eq{eq:Cmeasure}.

By using an identity $\delta g=g g^{\mu\nu} \delta g_{\mu\nu}$,
the shift of $C_{x_1,x_2,x_3}$ under an infinitesimal shift of the metric 
$\delta g_{\mu\nu}(x)$ is given by
\bea
\label{eq:dc}
\delta C_{x_1,x_2,x_3}&=&\Big[
\frac14 g^{\mu\nu}(x_1)\delta g_{\mu\nu}(x_1) +
\frac14 g^{\mu\nu}(x_2)\delta g_{\mu\nu}(x_2)+
\frac14 g^{\mu\nu}(x_3)\delta g_{\mu\nu}(x_3) \cr
&&\ -\beta
\delta \left( d(x_1,x_2)^2+d(x_2,x_3)^2+d(x_3,x_1)^2\right)
\Big]
C_{x_1,x_2,x_3}.
\eea
Putting \eq{eq:dc} into \eq{eq:Cmeasure}, one obtains
\bea
ds_g^2&=&B^2 \int d^Dx_1d^Dx_2d^Dx_3 \sqrt{g(x_1)}\sqrt{g(x_2)}\sqrt{g(x_3)} 
\Big[
\frac14 g^{\mu\nu}(x_1)\delta g_{\mu\nu}(x_1) +
\frac14 g^{\mu\nu}(x_2)\delta g_{\mu\nu}(x_2)\cr
&&\ \ \ \ \ +\frac14 g^{\mu\nu}(x_3)\delta g_{\mu\nu}(x_3) 
-\beta
\delta \left( d(x_1,x_2)^2+d(x_2,x_3)^2+d(x_3,x_1)^2\right)
\Big]^2 \cr
&&\hskip2cm \times \exp\left(-2 \beta
\left( d(x_1,x_2)^2+d(x_2,x_3)^2+d(x_3,x_1)^2\right)
\right).
\eea

Because of the last exponential damping factor, the integration over $x_2,x_3$
is dominated by the region within the distance of order $\sqrt{1/\beta}$ 
from $x_1$. 
Therefore if $g_{\mu\nu}(x)$ varies so slowly that 
the variation can be neglected in the length scale $\sqrt{1/\beta}$,
the values of $g_{\mu\nu}(x)$ in the integrand 
can be well approximated by $g_{\mu\nu}(x_1)$. 
This approximation will be available for the analysis of the 
low-momentum modes around a flat background, and 
the systematic improvement of this approximation could be obtained by derivative 
corrections with the expansion parameter $\sqrt{1/\beta}$.
Thus one obtains the corresponding metric in the space of $g_{\mu\nu}(x)$ 
in this slow varying approximation as 
\bea
\label{eq:dsslow1}
ds^2_{slow}&=&
B^2
\int d^Dx_1d^Dx_2d^Dx_3\, g(x_1)^{\frac32}\cr
&&\hskip.5cm\times \Big[
\frac34 g^{\mu\nu}(x_1)\delta g_{\mu\nu}(x_1)-\beta 
\delta \left( \tilde d(x_1-x_2)^2+\tilde d(x_2-x_3)^2+\tilde d(x_3-x_1)^2\right)
\Big]^2\cr
&&\hskip.5cm\times
\exp\left(-2 \beta
\left( \tilde d(x_1-x_2)^2+\tilde d(x_2-x_3)^2+\tilde d(x_3-x_1)^2\right)
\right),
\eea
where $\tilde d(x)$ is the distance defined with the metric $g_{\mu\nu}(x_1)$,
\be
\tilde d(x)^2\equiv g_{\mu\nu}(x_1)\, x^\mu x^\nu.
\ee
After expanding the square in the second line of \eq{eq:dsslow1} and 
rewriting, one obtains
\bea
\label{eq:ds2slow}
ds^2_{slow}&=&
B^2
\int d^Dx_1d^Dx_2d^Dx_3\, g(x_1)^{\frac32}\cr
&&\hskip.5cm\times \Bigg[
\left(\frac34\right)^2 \left(g^{\mu\nu}(x_1)\delta g_{\mu\nu}(x_1)\right)^2-
\frac{3\beta}{2} g^{\mu\nu}(x_1)\delta g_{\mu\nu}(x_1)\left(-\frac{1}{2\beta}\right)
\delta g_{\rho\sigma}(x_1) \frac{\partial}{\partial g_{\rho\sigma}(x_1)}\cr
&&\hskip1cm+\beta^2 \left(-\frac{1}{2\beta}\right)^2 
\delta g_{\mu\nu}(x_1)\delta g_{\rho\sigma}(x_1) 
\frac{\partial}{\partial g_{\mu\nu}(x_1)}
\frac{\partial}{\partial g_{\rho\sigma}(x_1)}
\Bigg]\cr
&&\hskip.5cm\times
\exp\left(-2 \beta
\left( \tilde d(x_2)^2+\tilde d(x_2-x_3)^2+\tilde d(x_3)^2\right)
\right)\cr
&=&
16 s_0
\int d^Dx\, g(x)^{\frac32} \Bigg[
\left(\frac34\right)^2 \left(g^{\mu\nu}(x)\delta g_{\mu\nu}(x)\right)^2+
\frac{3}{4} g^{\mu\nu}(x)\delta g_{\mu\nu}(x)
\delta g_{\rho\sigma}(x) \frac{\partial}{\partial g_{\rho\sigma}(x)}\cr
&&\hskip3cm+ \frac{1}{4} 
\delta g_{\mu\nu}(x)\delta g_{\rho\sigma}(x) 
\frac{\partial}{\partial g_{\mu\nu}(x)}
\frac{\partial}{\partial g_{\rho\sigma}(x)}
\Bigg]g(x)^{-1}\cr
&=& s_0 \int d^Dx \sqrt{g(x)} \left[
 \left(g^{\mu\nu}(x)\delta g_{\mu\nu}(x)\right)^2
+4 g^{\mu\rho}(x)g^{\nu\sigma}(x) \delta g_{\mu\nu}(x) 
\delta g_{\rho\sigma}(x)\right]. 
\eea
Here, from the first to the second equations, 
the Gaussian integration over $x_2,x_3$ has been performed 
and a numerical factor has been incorporated into a new numerical coefficient 
$s_0$,
and from the second to the last equations, I have used 
\bea
\frac{\partial}{\partial g_{\mu\nu}}g&=&g\, g^{\mu\nu}, \cr
\frac{\partial}{\partial g_{\mu\nu}}g^{\rho\sigma}&=&-g^{\rho\mu}g^{\nu\sigma}.
\eea
Thus the final expression  of $ds^2_{slow}$ \eq{eq:ds2slow} has the form of 
the DeWitt supermetric \cite{DeWitt:1962ud}, and its parameter has been fixed.

\subsection{Prediction of profiles of low-lying modes in $D=2$}
\label{sec:prediction}
In this subsection, I restrict the discussions to the small fluctuations 
around the $D=2$ flat background $g_{\mu\nu}=\delta_{\mu\nu}$.
I will first obtain the mode of the metric tensor 
transverse to the gauge directions \eq{eq:ggauge}, and
will then use the correspondence \eq{eq:correspondence} to obtain the corresponding
mode in the tensor model.  

Since the gauge transformation \eq{eq:ggauge} vanishes for a vanishing momentum,
all the components of the metric tensor survive at the zero-momentum sector.
As for a non-vanishing momentum, it is enough to 
consider the momentum $(p^1,0)\ (p^1\neq 0)$,
because of the rotational symmetry of the flat background.
Then the infinitesimal gauge transformation is given by
\bea
\label{eq:gaugedim2}
\delta g_{11}&=& 2i\, p^1\, v_1,\cr
\delta g_{22}&=& 0, \\
\delta g_{12}&=& i\, p^1\, v_2.\nonumber
\eea  
 
On the other hand, from \eq{eq:ds2slow}, the explicit form of 
the DeWitt supermetric in $D=2$ is given by
\be
\label{eq:explicitdewit}
ds^2_{slow}=s_0 \int d^2x \ 
(\delta g_{11},\delta g_{22},\delta g_{12})
\left(
\begin{array}{ccc}
5 & 1 & 0 \\
1 & 5 & 0 \\
0 & 0 & 8 
\end{array}
\right)
\left(
\begin{array}{c}
\delta g_{11} \\
\delta g_{22} \\
\delta g_{12}
\end{array}
\right).
\ee
Therefore the mode transverse to the gauge transformation \eq{eq:gaugedim2}
is given by    
\be
\left(
\begin{array}{c}
\delta g_{11} \\
\delta g_{22} \\
\delta g_{12}
\end{array}
\right)
\propto \left(
\begin{array}{ccc}
5 & 1 & 0 \\
1 & 5 & 0 \\
0 & 0 & 8 
\end{array}
\right)^{-1} 
\left(
\begin{array}{c}
0 \\
1 \\
0
\end{array}
\right)
=\frac{1}{24}
\left(
\begin{array}{c}
-1 \\
5 \\
0
\end{array}
\right).
\ee
Thus the transverse mode is characterized by
\bea
\label{eq:tranmode}
\frac{\delta g_{11}}{\delta g_{22}}&=&-\frac15,\cr
\frac{\delta g_{12}}{\delta g_{22}}&=&0.
\eea

The corresponding transverse mode in the tensor model can be obtained by
putting \eq{eq:tranmode} into the correspondence \eq{eq:correspondence}.
One may directly compare the obtained mode with the low-lying modes from 
the numerical analysis.
However, the direct comparison through $C_{abc}$ is inconvenient, because
it has rather many components and its dependence on the metric tensor is rather 
complicated and is difficult to see\footnote{In fact, I have performed the 
direct comparison in some cases, and have obtained the same results.}. 
In fact,
it is much more convenient to use the two-tensor $K_{ab}$ defined in
\eq{eq:defk}.
Under the correspondence \eq{eq:correspondence}, the dependence of the two-tensor
on the metric tensor can be obtained as
\bea
K_{x_1, x_1'}&=&
\int d^Dx_2 d^Dx_3\,
C_{x_1,x_2,x_3} C_{x_1',x_2,x_3} \cr
&=&
B^2 
g(x_1)^\frac{1}{4} g(x_1')^\frac{1}{4}
\int d^Dx_2 d^Dx_3
\sqrt{g(x_2)}\sqrt{g(x_3)} \cr
&& 
\times\exp
\left[
-\beta \left(d(x_1,x_2)^2+d(x'_1,x_2)^2+d(x_1,x_3)^2+d(x_1',x_3)^2+2d(x_2,x_3)^2
\right)
\right] \cr
&=&
K_0\, g(x_1)^\frac14 g(x_1')^\frac14
\exp
\left(
-\beta\, d(x_1,x_1')^2
\right),
\eea
where I have used the slow varying approximation mentioned previously 
to integrate over $x_2,x_3$, and $K_0$ is a numerical factor. 

For a small perturbation from the flat background,
the variation of $K$ is given by
\bea
\label{eq:delk}
\delta K_{x_1,x_2}&=&
K_0 \left[ \frac12 \delta g_\mu{}^\mu\left(\frac{x_1+x_2}{2}\right)
-\beta\, \delta g_{\mu\nu}\left(\frac{x_1+x_2}{2}\right)(x_1-x_2)^\mu (x_1-x_2)^\nu
\right]\cr
&&\ \ \ \times\exp\left(-\beta (x_1-x_2)^2\right),
\eea
where the values of the metric tensor are replaced with
the representative value $g_{\mu\nu}((x_1+x_2)/2)$ by using the 
slow varying approximation, keeping the symmetry of $K_{x_1,x_2}$ under the exchange
of the indices.  
Since the numerical analysis will be performed in the momentum representation,
it is more convenient to have $\delta K$ in the momentum representation.  
Performing the Fourier transformation mentioned in Section~\ref{sec:themodel},
one obtains
\bea
\label{eq:dkp}
\delta K_{p_1,p_2}&=&F_{p_1}{}^{x_1} F_{p_2}{}^{x_2}\delta K_{x_1,x_2} \cr
&=&\frac{1}{(2\pi)^D}
\int d^Dx_1d^Dx_2\,
e^{ip_1x_1+ip_2x_2}\,
\delta K_{x_1,x_2} \cr
&=&K_p^0\,
\delta g_{\mu\nu}(p_1+p_2)\, 
(p_1-p_2)^\mu (p_1-p_2)^\nu \exp\left(-\frac1{16\beta}(p_1-p_2)^2 \right), 
\eea
where $K_p^0$ is a numerical factor and $\delta g_{\mu\nu}(p)$ is the Fourier
transform of $\delta g_{\mu\nu}(x)$.

When a fluctuation mode $\delta C_{abc}$ is obtained in numerical analysis, 
the variation of the two-tensor is given by
\be
\label{eq:tensordk}
\delta K_{ab}=\delta C_{acd}\, C^0{}_{b}{}^{cd}+  C^0_{acd}\, \delta C_{b}{}^{cd},
\ee  
where $C^0_{abc}$ is a classical solution. 
The formulas \eq{eq:tensordk} and \eq{eq:dkp} 
make comparable the fluctuation modes in tensor models and the 
metric modes in the general relativity.  

\begin{figure}
\begin{center}
\includegraphics[scale=.5]{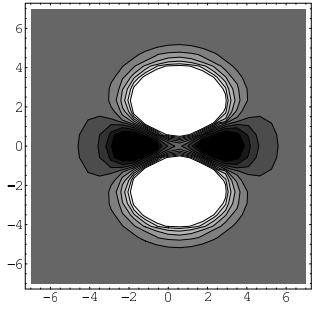}
\end{center}
\caption{The contour plot of $\delta K^{trans}(q)$ with $p^1=1,\ \beta=1$. 
The axes are $q=(q^1,q^2)$.
This choice of the parameters $p^1,\beta$ is just for an example,
and therefore only the qualitative characters of the figure is relevant.
 }
\label{fig:profile}
\end{figure}

It is worthwhile to see $\delta K$ for the transverse mode \eq{eq:tranmode}
at the momentum sector $(p^1,0)$ in two dimensions.
Putting \eq{eq:tranmode} into \eq{eq:dkp} and noting $p_1+p_2=(p^1,0)$ must be 
satisfied\footnote{In the continuum case, this is described by a delta function, and
this must be inserted into \eq{eq:ktrans}. But this is abbreviated for simplicity, since the actual interest is in tori, which have discrete momenta.}
for $\delta K$ in \eq{eq:dkp} to be non-vanishing,
the momentum dependence of $\delta K$ is obtained as 
\be
\label{eq:ktrans}
\delta  K^{trans}(q)\equiv\delta K_{q,-q+(p^1,0)}
=k_0
\left(\left(q^1-\frac{p^1}2\right)^2-5(q^2)^2\right)
\exp\left(-\frac{1}{4\beta}\left(
\left(q^1-\frac{p^1}2\right)^2+\left(q^2\right)^2\right)\right),
\ee
where $q=(q^1,q^2)$ and $k_0$ is a numerical factor. 
The profile of $\delta K^{trans}(q)$ is plotted in
Figure~\ref{fig:profile}.

\begin{figure}
\begin{center}
\includegraphics[scale=.5]{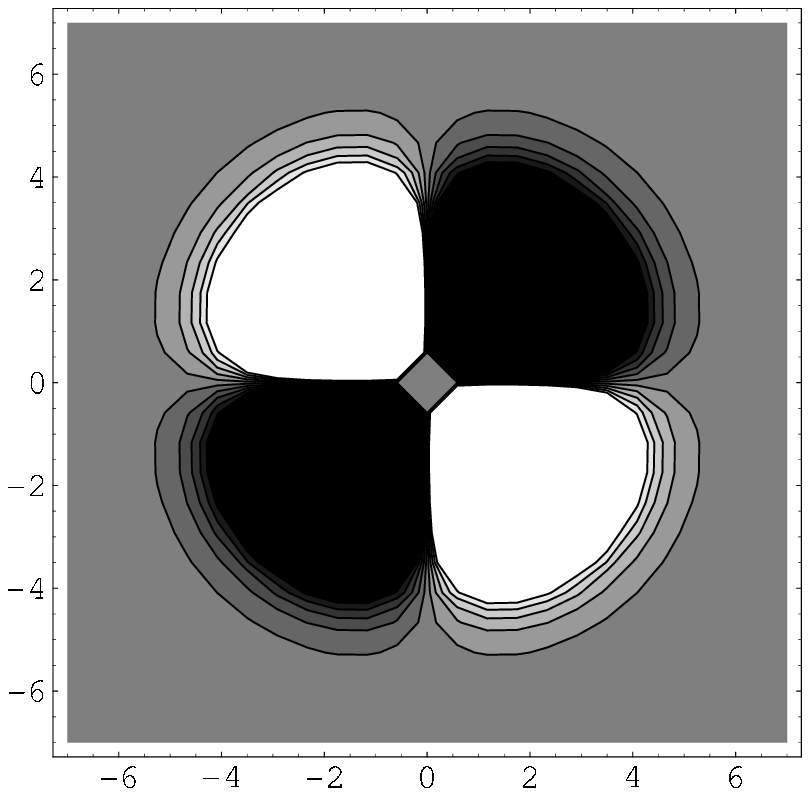}
\hfil
\includegraphics[scale=.5]{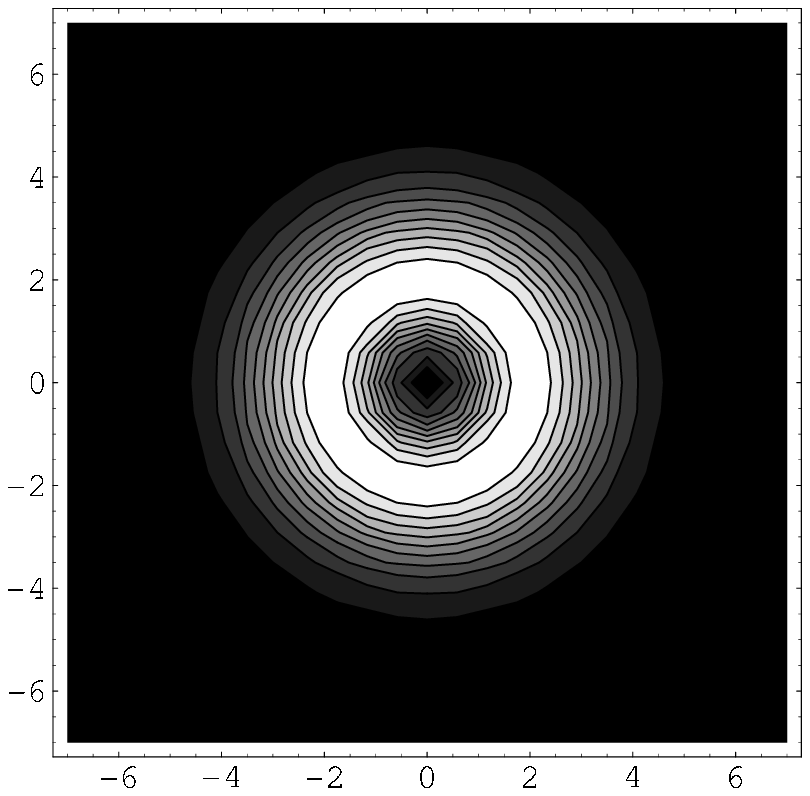}
\hfil
\includegraphics[scale=.5]{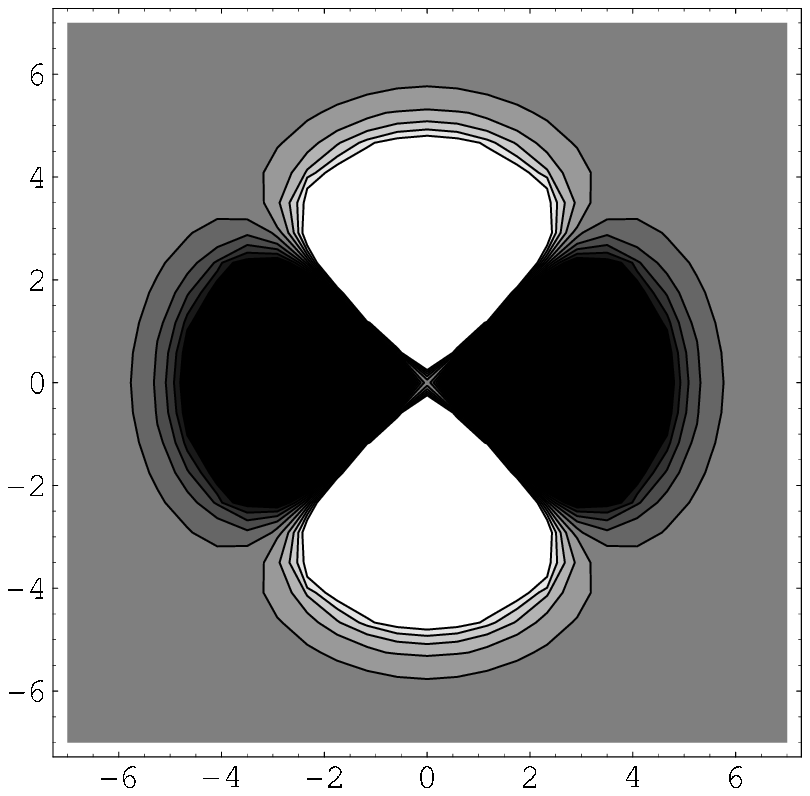}
\end{center}
\caption{The contour plots of $\delta K^{zero}(q)$ with $p=(0,0),\ \beta=1$
for the zero momentum fluctuations. The axes are $q=(q^1,q^2)$. 
From left to right, \eq{eq:threezero1},
\eq{eq:threezero2} and \eq{eq:threezero3}, respectively.
 }
\label{fig:profilezero}
\end{figure}

At the $p=(0,0)$ sector on a flat torus, 
as can be seen in \eq{eq:ggauge}, the gauge symmetry is null, and 
all the metric components survive. The DeWitt supermetric \eq{eq:explicitdewit}
implies that 
the eigenmodes will be given by the following three ones,
\bea
\label{eq:threezero1}
\delta g_{11}=\delta g_{22}=0,&& \delta g_{12}\neq 0,\\
\label{eq:threezero2}
\delta g_{11}=\delta g_{22}\neq 0,&& \delta g_{12}=0,\\
\label{eq:threezero3}
\delta g_{11}=-\delta g_{22}\neq 0,&& \delta g_{12}=0.
\eea
The corresponding $\delta K^{zero}(q)\equiv\delta K_{q,-q}$ are plotted 
in Figure~\ref{fig:profilezero}.

\section{Comarison with numerical analysis}
\label{sec:numerical}
The main topic of this section is to compare the properties of the low-lying
modes of the tensor model 
with the general relativity by using the correspondence \eq{eq:correspondence}.
Because of the simplicity of the fitting process, the actual comparison will be performed through $\delta K$, \eq{eq:dkp} and \eq{eq:tensordk}. 

Let me first review the method to obtain the spectra and the eigenmodes around a 
solution.
Let me assume an action $S(C)$ of a Euclidean tensor model 
has a classical solution $C^0_{abc}$,
\be
\left.
\frac{\partial S(C)}{\partial C_{abc}}\right|_{C=C^0}=0.
\ee 

The spectra of the quadratic potential for the fluctuations around a classical 
solution depend on the normalization of the fluctuations.
The natural normalization can be obtained from the measure of the path integral
in the space of $C_{abc}$.
The simplest $O(n)$ invariant measure is given by \eq{eq:Cmeasure}
assumed in the previous section. 
Since $C_{abc}$ is symmetric under the permutations of its indices,
the measure for its independent components is given by
\be
\label{eq:Cmeasuresym}
ds^2_C=\delta C_{abc}\, \delta C^{abc}=\sum_{(a,b,c)} m[(a,b,c)]\, 
\delta C_{(a,b,c)}\, \delta C^{(a,b,c)},
\ee
where $(a,b,c)$ denotes a set of $a,b,c$ independent of their order,
$C_{(a,b,c)}=C_{abc}$, 
and $m[(a,b,c)]$ is the multiplicity defiend by
\be
\label{eq:multifac}
m[(a,b,c)]=\left\{
\begin{array}{cl}
1 & a=b=c, \\
3 & a=b\neq c,\ b=c\neq a, {\rm \ or\ } c=a \neq b,\\
6 & {\rm otherwise}.
\end{array}
\right.
\ee
Therefore the normalized components can be given by
\be
\delta \tilde C_{(a,b,c)}=\sqrt{m[(a,b,c)]}\delta C_{(a,b,c)},
\ee 
and the coefficient matrix of the quadratic potential for these components is
\bea
\label{eq:defm}
M^{(a,b,c),(d,e,f)}&=&\left. \frac12 
\frac{\partial^2 S(C)}{\partial \tilde C_{(a,b,c)} \partial \tilde C_{(d,e,f)}}
\right|_{C=C^0} \cr
&=&
\left. \frac12 
\frac{1}{\sqrt{m[(a,b,c)]m[(d,e,f)]}}
\frac{\partial^2 S(C)}{\partial C_{(a,b,c)} \partial C_{(d,e,f)}}
\right|_{C=C^0} .
\eea

The spectra and the eigenmodes of the quadratic potential can be obtained by
diagonalizing the symmetric matrix \eq{eq:defm}.
The spectra contain a number of zero modes, which 
come from the symmetry breaking of the original $O(n)$ symmetry of the Euclidean
tensor model to the remaining symmetry of a classical solution $C^0$.
According to the idea of \cite{Borisov:1974bn}, 
these zero modes should be identified as 
the gauge symmetry non-linearly realized on a certain background $C^0$.
Because the eigenmodes of a symmetric matrix like \eq{eq:defm} 
are transverse to each other, the ``physical" modes
reside in the subspace transverse to the space of the zero modes.  
This justifies the reason why special attention is paid to 
the transverse mode to the general coordinate transformation in 
Section~\ref{sec:prediction}. 

The numerical analysis is performed on a Windows XP64 workstation
with two Opteron 275 (2.2GHz, dual core each) processors and 8 GB memories. 
The C++ codes are
compiled by the Intel C++ compiler 10.0 with OMP parallelization. 
NAG C Library Mark 8 and ACML 3.6.0 \& 4.0.0 are used for numerical routines. 
Mathematica 5.2 is used for analyzing the outputs.

\subsection{Fuzzy flat tori}
\label{sec:flattori}
The classical solutions which will be 
considered in this subsection are Gaussian-like numerical 
solutions representing
fuzzy flat tori analogous to the analytic Gaussian solutions \eq{eq:cx} 
or \eq{eq:cp} for fuzzy flat spaces. The strategy of the numerical 
analysis is basically the same as the previous paper \cite{Sasakura:2007sv}. 
I assume such a solution for a fuzzy flat torus has an 
$SO(2)\times SO(2)$ remaining symmetry in the same way as 
the translational symmetry of a usual flat torus.
Because of the symmetry, momentum is a conserved quantity so that
it is convenient to use the momentum representation in the numerical analysis.
The momentum takes discrete values because of the finite size of the torus.
Thus two-dimensional discrete momentum vectors are taken as the tensor index,
\be
p=(p^1,p^2),\ \ (p^i=-L,-L+1,\cdots,L),
\ee
where $L$ is a UV cutoff and is a positive integer.  
Because of the momentum conservation, a classical solution can be assumed to 
take a form,
\bea
C^0_{p_1,p_2,p_3}&=&\delta_{p_1+p_2+p_3,(0,0)}\, A(p_1,p_2,p_3), \\
g^{p_1,p_2}&=&\delta_{p_1+p_2,(0,0)},
\eea
where $A(p_1,p_2,p_3)$ are numerical coefficients defined only for $p_1+p_2+p_3=(0,0)$,
and symmetric for the momentum variables.
The measure for contracting indices is just the discrete sum $\sum_p$.

As was stressed in Section~\ref{sec:themodel}, the momentum representation 
is just a technically convenient representation, and to embed the solution
into the Euclidean tensor model, one has to consider a coordinate representation.
The discrete form of the Fourier transformation can be given by the 
transformation matrix, 
\bea
F_x{}^p&=&\frac{1}{2L+1}\,\exp\left(\frac{2\pi i}{2L+1}\, p\, x \right),\\
\label{eq:discfourierpx}
F_p{}^x&=&\frac{1}{2L+1}\,\exp\left(-\frac{2\pi i}{2L+1}\, p\, x \right),
\eea
where the coordintate $x$ takes finite discrete values,
\be
\label{eq:disccoord}
x=(x^1,x^2),\ \ \left(x^i=-L,-L+1,\cdots,L \right),
\ee
and $F_x{}^p$ and $F_p{}^x$ are inverse to each other.
The relation between the tensors in the momentum and the coordinate representations
is given by 
\bea
\label{eq:discx}
C_{x_1,x_2,x_3}&=&F_{x_1}{}^{p_1}F_{x_2}{}^{p_2}F_{x_3}{}^{p_3}C_{p_1,p_2,p_3}, \\
\label{eq:disgx}
g^{x_1,x_2}&=&F_{p_1}{}^{x_1}F_{p_2}{}^{x_2}g^{p_1,p_2}=\delta_{x_1,x_2}.
\eea
Because of \eq{eq:disgx}, the solution can now be embedded into a Euclidean tensor model, if $C_{x_1,x_2,x_3}$ in \eq{eq:discx} is real. This is satisfied,
if
\be
A(p_1,p_2,p_3)^*=A(-p_1,-p_2,-p_3).
\ee

In the actual process of searching for a solution, I have assumed two more ansatz.
One is that $A(p_1,p_2,p_3)$ is real. The other is that $A(p_1,p_2,p_3)$ is 
invariant under the following three reflection symmetries,
\bea
(p_i^1,p_i^2)&\rightarrow &(-p_i^1,p_i^2), \\
(p_i^1,p_i^2)&\rightarrow &(p_i^1,-p_i^2), \\
(p_i^1,p_i^2)&\rightarrow &(p_i^2,p_i^1), 
\eea
For example, $A((p_1^1,p_1^2),(p_2^1,p_2^2),(p_3^1,p_3^2))=
A((-p_1^1,p_1^2),(-p_2^1,p_2^2),(-p_3^1,p_3^2))$, etc.
These ansatz reduce the number of free variables, and simplify the 
process of solution search. 
These ansatz mean that a solution describes a fuzzy analogue of $S^1 \times S^1$,
where the two $S^1$'s have the same size and a reflection symmetry, 
and are transverse to each other.

Stable classical solutions to the action $S(C)$ in \eq{eq:theaction}
can be found by searching for its local minima.
To reduce the free variables, the above ansatz are put into $S(C)$ to make $S(A)$, 
and local minima of $S(A)$ are numerically searched by a NAG routine.
Actually various solutions to \eq{eq:generaleq} ($S(A)$=0) has been found.
Since $S(C)$ is a semi-positive definite function of $C$, these solutions 
are actually solutions to $S(C)$ with respect to $C$, 
and all the fluctuation spectra around it are non-negative.

\begin{figure}
\begin{center}
\includegraphics[scale=.7]{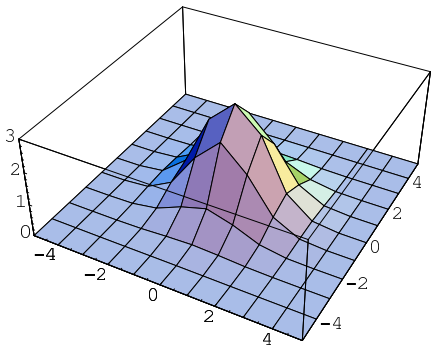}
\end{center}
\caption{The plot of $K_{p,-p}$ for a Gaussian-like solution for $L=5$.
The horizontal axes are $p=(p^1,p^2)$.  }
\label{fig:kl5}
\end{figure}

For example, one can find solutions very similar to the Gaussian solutions
\eq{eq:cp}. In Figure~\ref{fig:kl5}, the tensor $K_{p,-p}$ 
in \eq{eq:defk} computed for a solution for $L=5$ is plotted. The profile has a 
Gaussian-like form.  

Since momentum is the conserved quantity of the classical backgrounds, the 
basic strategy to obtain the fluctuation spectra and mode profiles
around the solutions is to obtain the eigenvalues and eigenvectors of the
submatrix in each momentum sector of the matrix \eq{eq:defm}.
The technical details are given in Appendix \ref{sec:app}.  
 
As same as what was obtained in the previous paper \cite{Sasakura:2007sv}, 
the spectra can be classified into three categories. 
One is the category of zero-modes, another is the low-lying modes, and 
the other is the other ``heavy" modes.

Let me first review the zero modes discussed in detail in the 
previous paper \cite{Sasakura:2007sv}. 
The zero modes come from the symmetry breaking 
of the $O(n)\ (n=(2L+1)^2)$ symmetry to $SO(2)\times SO(2)$ by the classical solutions.
Following the idea of \cite{Borisov:1974bn}, 
these zero modes should be regarded as the gauge symmetry
non-linearly realized on a background.
Since the $O(n,R)$ transformation operates on $C_{x_1,x_2,x_3}$ in the coordinate representation,
the generators are characterized by
\bea
M_{x_1,x_2}{}^*&=&M_{x_1,x_2}, \cr
M_{x_1,x_2}&=& -M_{x_2,x_1}, 
\eea 
where $x_i$ are the discrete coordinates \eq{eq:disccoord}.
In the momentum representation, after the Fourier transformation $F_{p}{}^x$ in
\eq{eq:discfourierpx}, one 
obtains the corresponding conditions as
\bea
M_{p_1,p_2}{}^*&=&M_{-p_1,-p_2}, \cr
M_{p_1,p_2}&=&-M_{p_2,p_1}.
\eea 
The number counting of these generators at each momentum sector was 
performed in the previous paper \cite{Sasakura:2007sv}. 
Taking into account the existence of 
the unbroken generators in the $p=(0,0)$ sector,
one obtains the formula for the number of the zero modes 
at each momentum sector for $D=2$ as
\be
\label{eq:numzero}
\# {\rm zero}(p)=
\left\{ 
\begin{array}{ll}
\frac12 \left( \prod_{i=1}^2 \left( 2L+1-|p^i|\right) -{\rm even}(p)\right)
-2 \delta_{p,(0,0)} &{\rm for\ } |p^i|\leq 2L,\\
0&{\rm otherwise}, 
\end{array}
\right.
\ee
where ${\rm even}(p)$ denotes a function of whether $p$ is an even vector $p=(2i,2j)$
with $i,j$ integers or not,
\be
{\rm even}(p)=\left\{
\begin{array}{cl}
1 & {\rm for\ an\ even\ vector\ }p, \\
0 & {\rm otherwise}.
\end{array}
\right.
\ee
One can check that the number of very tiny eigenvalues in the order of  
machine errors $\lesssim 10^{-15}$ at each momentum sector agrees with
the formula \eq{eq:numzero}. 

\begin{figure}
\begin{center}
\includegraphics[scale=.7]{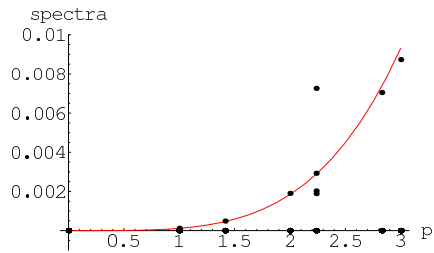}
\end{center}
\caption{The spectra of the quadratic fluctuations around the $L=5$ solution.
 The horizontal axis is $|p|=\sqrt{(p^1)^2+(p^2)^2}$. The solid line is
$0.000115 |p|^4$. At the $p=(0,0)$ sector,
there exist three extra spectra other than the zero modes, 
although they cannot be distinguished from the origin in the figure. 
At each $|p|\neq 0$ sector except $p=(2,1)\ ({\rm and}\ (1,2),\ {\rm e.t.c.})$,
there exists one low-lying spectrum.}
\label{fig:specL5}
\end{figure}

The low-lying modes are the modes which are expected to describe the ``low-energy"
effective dynamics of the system, and will be identified with the metric modes
of the general relativity in the sequel. 
In Figure~\ref{fig:specL5}, the spectra of the quadratic
fluctuations at some low-momentum sectors around the $L=5$ solution above are shown.  
As was obtained in the previous paper \cite{Sasakura:2007sv},
there exist three low-lying modes at the $p=(0,0)$ sector, and one at each 
non-zero momentum sector except $p=(2,1)\ ({\rm and}\ (1,2),\ {\rm e.t.c.})$.
At this sector, 
some of the ``heavy" modes seem to overlap with the low-lying mode. This
would be an accidental phenomenon caused by the smallness of $L$, 
and may be improved for 
larger $L$. 
Rather than going to larger $L$, which requires much larger machine power, 
 a systematic renormalization-like procedure 
will be discussed in Section~\ref{sec:clasren}, and the spectral pattern of the low-lying modes 
will become more evident. 
The low-lying modes seem to be well on the line of a massless trajectory with 
the momentum dependence of $|p|^4$.  Any other momentum 
dependence such as $|p|^2$ does not fit well with the data.

Since the Einstein-Hilbert action $\int d^2x \sqrt{g} R$ is a topological term 
in two dimensions, the lowest momentum dependence of the transverse modes 
will come from the curvature square term. 
Therefore the quartic momentum dependence of
the trajectory is natural at least in two dimensions.

\begin{figure}
\begin{center}
\includegraphics[scale=.5]{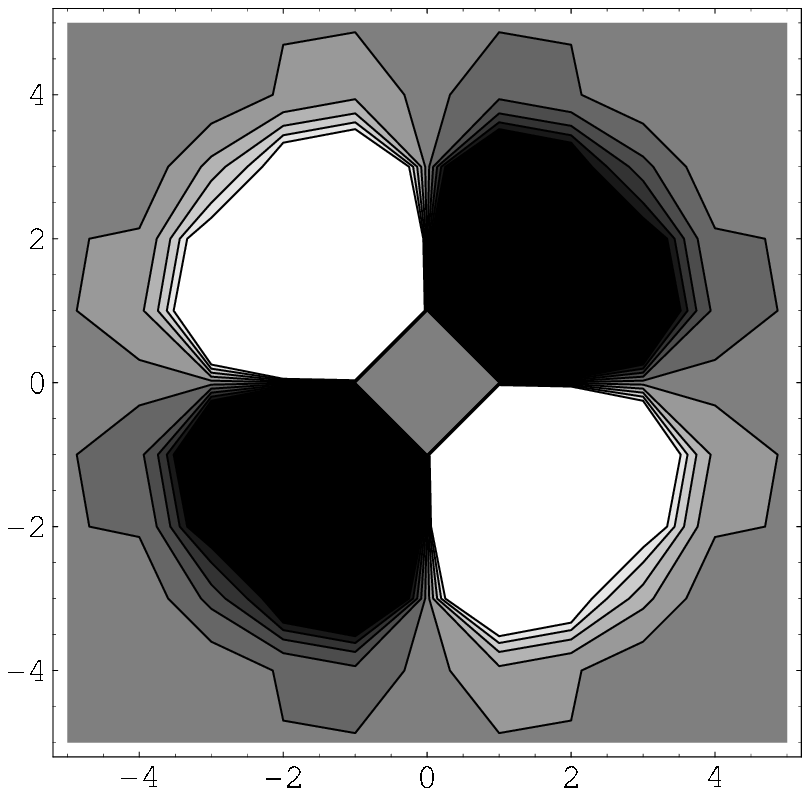}
\hfil
\includegraphics[scale=.5]{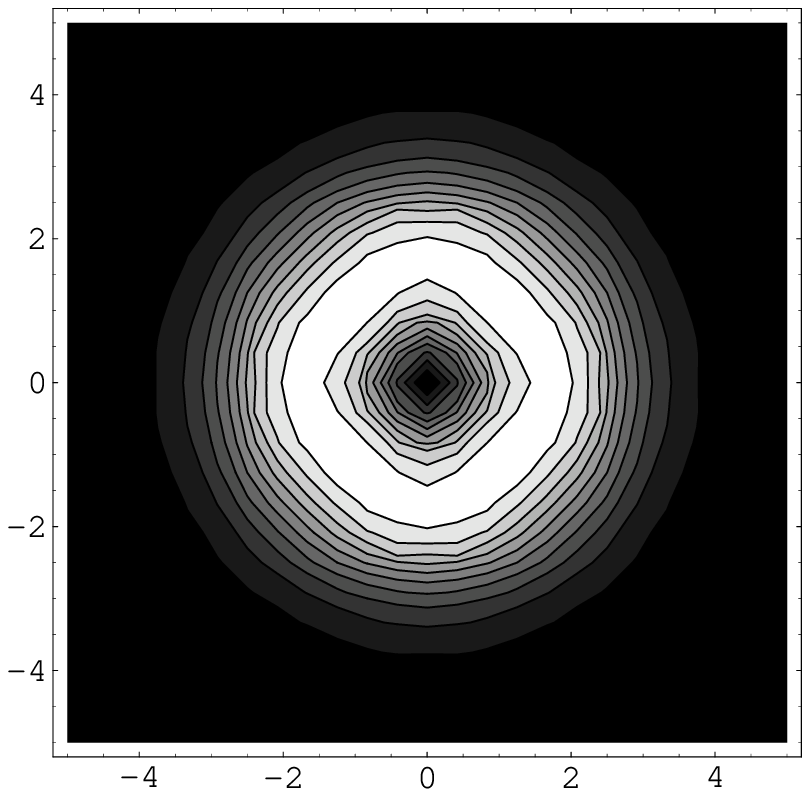}
\hfil
\includegraphics[scale=.5]{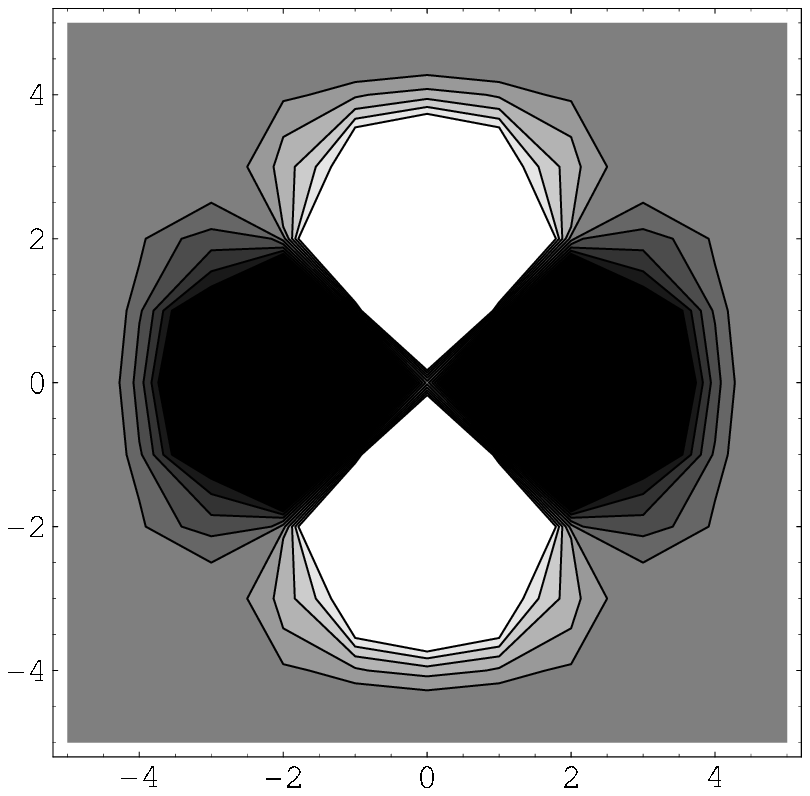}
\end{center}
\caption{The contour plots of $\delta K_{q,-q}$ 
for the three low-lying modes at the $p=(0,0)$ sector of the $L=5$ solution. 
The axes are $q=(q^1,q^2)$.
From left to right, the spectral values of the modes are $10^{-15}$, 
which is the order of 
the machine errors, $4\times 10^{-10}$ and $8\times 10^{-10}$, respectively.
The figures are in good qualitative agreement with Figure~\ref{fig:profilezero}, which 
correspond to the metric fluctuations \eq{eq:threezero1}, \eq{eq:threezero2} and 
\eq{eq:threezero3}.
 }
\label{fig:threezerotens}
\end{figure}

The eigenvectors of each spectra determines the fluctuations $\delta C_{abc}$ 
of each mode. 
Putting these into \eq{eq:tensordk}, one can compute the corresponding $\delta K_{ab}$.
These are drawn for the three low-lying modes at the $p=(0,0)$ sector for the 
$L=5$ solution in Figure~\ref{fig:threezerotens}. 
Comparing with Figure~\ref{fig:profilezero},
the ``lightest" mode can be identified with the metric mode \eq{eq:threezero1},
which changes the relative angle of the two $S^1$'s.
The second one is \eq{eq:threezero2}, which is the change of the whole size. 
The last is \eq{eq:threezero3}, which changes the relative sizes of the two $S^1$'s.

\begin{figure}
\begin{center}
\includegraphics[scale=.5]{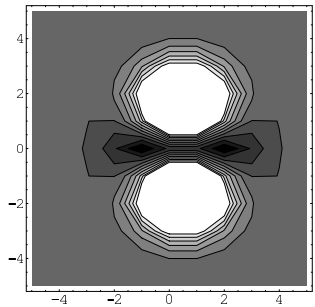}
\end{center}
\caption{The contour plot of $\delta K_{q,-q+(1,0)}$ 
for the low-lying mode at the $p=(1,0)$ sector of the $L=5$ solution. 
The axes are $q=(q^1,q^2)$.
Its spectrum is $1.2\times 10^{-4}$.
The figure agrees qualitatively very well with Figure~\ref{fig:profile}.}
\label{fig:p1}
\end{figure}

Similarly, the $\delta K_{ab}$ for the low-lying mode at the $p=(1,0)$ sector 
is shown in Figure~\ref{fig:p1}.
One can see that this looks very similar to $\delta K^{trans}(q)$ in 
Figure~\ref{fig:profile}, which
corresponds to the metric fluctuation transverse to the general coordinate transformation.
From a statistical analysis, the parameter fitting with the formula \eq{eq:dkp} 
results in $\delta g_{11}/\delta g_{22} =-0.18\pm 0.01,\ \delta g_{12}\simeq 0$.
This is in agreement with \eq{eq:tranmode}.

These evidences show clear matching between the low-lying modes in the tensor model
and the transverse metric fluctuations in the general relativity.

\section{Renormalization procedure for tensor models}
\label{sec:ren}
In this section, I will discuss a Wilson's type of 
renormalization procedure in tensor models.
A standard application of such a renormalization procedure is to
study quantum dynamics of field theory 
by considering effective actions of renormalized fields. 
The study of quantum properties of tensor models is surely an important direction, 
but is out of the scope of the present work.
In this section, I will rather use the procedure as a systematic method to single out
slow-varying modes.

\subsection{A proposal for renormalization procedure}
\label{sec:renproc}
The general philosophy of the Wilson's type of renormalization procedures 
is in the process of coarse graining. 
High frequency (momentum) modes over a cutoff 
are integrated out to produce an effective field theory of low-frequency modes below the cutoff. The non-trivial issue in the gravitational case is that, since
the scale itself is dynamical, one cannot set a cutoff independently from 
dynamical variables.

In a discrete system, the procedure is generally a discrete step of defining a 
new renormalized field by averaging over some nearby site contributions.   
In the application to tensor models, since a cutoff itself must be determined from the 
tensor $C_{abc}$ itself as a gravitational system, 
the procedure would be described by reputation of a step described by
\be
C^{(r)}_{abc}=R\left( C^{(r-1)}\right)_{abc},
\ee
where $R$ is a function of the tensor $C_{abc}$, 
and $r=0,1,\cdots$ labels the number of the steps.
Since the $O(n)$ symmetry is an essential ingredient of tensor models
and it is hard to control its breaking, 
the function $R$ must keep the invariance. 
  One of the simplest choices of such $R$ is given by the $H_{abc}$ in \eq{eq:defk},  
\be
\label{eq:renc}
C^{(r)}_{abc}=R\left(C^{(r-1)}\right)_{abc}=C^{(r-1)}{}_{ad}{}^e C^{(r-1)}{}_{be}{}^f C^{(r-1)}{}_{cf}{}^d.
\ee

To check that this is a kind of coarse graining process, let me consider \eq{eq:cp} 
(or \eq{eq:cx}) as an input. 
From \eq{eq:barh}, one can see that the Gaussian form of \eq{eq:cp} is kept under the 
renormalization process, and obtains a discrete renormalization flow, 
\be
\label{eq:renalpha}
\alpha^{(r)}=\frac{5}{3} \alpha^{(r-1)},
\ee
where $\alpha^{(r)}$ denotes the parameter $\alpha$ in \eq{eq:cp} for the $r$-th 
renormalized tensor $C^{(r)}$. 
After one step, the value of $\alpha$ increases, and
the fuzziness of the space becomes larger, because the algebra defined by 
$f_{x_1}f_{x_2}=C^{(r)}_{x_1x_2}{}^{x_3}f_{x_3}$ becomes more widely spread
than that determined by $C^{(r-1)}$. 
Therefore the effect caused by the renormalization step can be regarded as a 
kind of coarse graining process averaging over the contributions in a nearby 
region.

\subsection{Classical application and trajectory of spectra}
\label{sec:clasren}
The problem I consider in this subsection is the overlapping of the ``heavy" modes
on the low-lying modes encountered at the $p=(2,1)$ sector 
in the $L=5$ spectra as in Figure~\ref{fig:specL5}.
For this small size of $L$, it is difficult to judge whether the overlapping ``heavy"
modes are relevant in the large $L$ limit or not. In general, a renormalization
procedure should improve this kind of ambiguity, since a renormalized system  
has a larger effective size. 
Therefore it would be interesting to
see how the renormalization procedure discussed in the previous 
subsection affects the spectra. 

In the path integral formulation (Euclidean), 
the formal strategy to obtain an effective action
is given by
\bea
\int {\cal D}C^{(r-1)}\, e^{-S^{(r-1)}\left( C^{(r-1)} \right)}
&=&\int  {\cal D}C^{(r)}{\cal D}C^{(r-1)}\, \delta\left(C^{(r)}-R\left(C^{(r-1)}\right)\right)\, e^{-S^{(r-1)}\left( C^{(r-1)} \right)}\cr
&=& \int  {\cal D}C^{(r)}\, e^{-S^{(r)}\left( C^{(r)} \right)},
\eea
where 
\be
e^{-S^{(r)}\left( C^{(r)} \right)}
=\int {\cal D}C^{(r-1)}\, \delta\left(C^{(r)}-R\left(C^{(r-1)}\right)\right)
\, e^{-S^{(r-1)}\left( C^{(r-1)} \right)}.
\ee
Therefore the effective action is obtained by
\be
\label{eq:renS}
S^{(r)}\left(C^{(r)}\right)=S^{(r-1)}\left(R^{-1}\left(C^{(r)}\right)\right)+
\Delta^R\left(C^{(r)}\right),
\ee
where $\Delta^R$ is the contribution from the determinant associated to the
change of variables.
 
In the following I will only consider the classical application of the renormalization
procedure, namely, the first term of \eq{eq:renS}, and leave the analysis of
the full contributions for future study.
I also consider only $S^{(1)}$, which is obtained by a one step of renormalization
from the original one.

After one step of the classical renormalization, 
the equation of motion \eq{eq:generaleq} is given by
\be
\label{eq:eomren}
W^{(1)}_{abc}\left(C^{(1)}\right)=W_{abc}\left(R^{-1}\left(C^{(1)}\right)\right)=0.
\ee
Therefore a solution to $W^{(1)}_{abc}=0$ is just the renormalization of the 
original solution $C^0$,
\be
C^{(1)\,0}=R(C^0).
\ee
Putting \eq{eq:eomren} into the first derivative 
$\partial W^{(1)}_{abc}/\partial C^{(1)}_{def}$ at the solution, one obtains
\bea
\label{eq:w1c}
\left.
\frac{\partial W^{(1)}_{abc}}{\partial C^{(1)}_{def}}\right|_{C^{(1)}=R^{-1}(C^0)}=
\left.
\frac{\partial W_{abc}}{\partial C_{ghi}}
\frac{\partial C_{ghi}}{\partial C^{(1)}_{def}}
\right|_{C=C^0}.
\eea
As in Appendix~\ref{sec:app},
the spectra of the quadratic fluctuations can basically be obtained from
the square of the matrix \eq{eq:w1c} with inclusion of the multiplicity factors
\eq{eq:multifac}.

Although the above is the theoretically correct procedure for obtaining the spectra, 
a serious difficulty appears in the actual numerical computation.
That is, the NAG and ACML routines for eigenvalues cannot produce reliable values.
This can be checked by comparing the obtained eigenvalues with the formula
\eq{eq:numzero} of the number of the zero modes. 
The direct reason for this malfunction is that the matrix 
$\partial C_{abc}/\partial C^{(1)}_{def}$
in \eq{eq:w1c}  contains very large components and enhances the 
numerical errors.
The intuitive reason for such large components can be given as follows.
The renormalization step will  pick out slow-varying important components, 
but suppress rapidly-changing unimportant ones. 
Therefore the derivative $\partial C^{(1)}_{abc}/\partial C_{def}$ will generally 
be very small for unimportant components of $C_{def}$. 
These small values of components will lead to a number of large components of 
$\partial C_{abc}/\partial C^{(1)}_{def}$, which is 
the inverse of the matrix $\partial C^{(1)}_{abc}/\partial C_{def}$.
  
To overcome the problem, let me use the fact that the matrix $M^{(1)}$
of the quadratic fluctuations can be put into a form, 
\be
M^{(1)}=H^T\, H,
\ee
where $H$ is a matrix and $T$ denotes the transpose,
since $M^{(1)}$ is essentially a square of \eq{eq:w1c} as in Appendix~\ref{sec:app}. 
Even though $H$ contains large values,
the computation with $H$ turns out to be much better than directly dealing 
with $M^{(1)}$.
To compute a few of the smallest eigenvalues over the zero modes, 
successive minimum searches of
\be
\frac{(H x_\perp)^T\, H x_\perp  }{x_\perp^T\, x_\perp}
\ee
have been performed. Here $x_\perp$ is the variable real vector, 
and varies in the vector 
space transverse to the zero modes discussed in Section~\ref{sec:flattori}, 
when the lowest non-zero eigenvalue is searched. 
When the second lowest is searched,
the $x_\perp$ varies in the vector space transverse to the lowest non-zero 
mode as well. This iterative procedure can be continued until a satisfactory number of
the lowest eigenvalues are obtained. 
The inner product comes from the measure \eq{eq:Cmeasuresym}.

\begin{figure}
\begin{center}
\includegraphics[scale=.7]{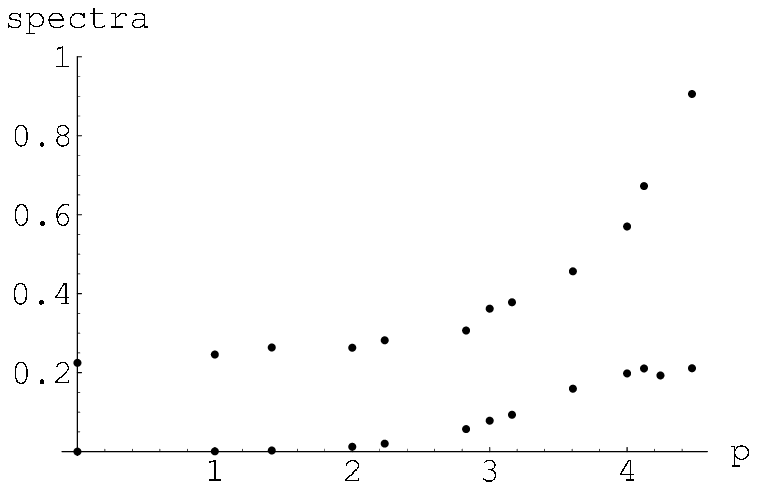}
\hfil
\includegraphics[scale=.7]{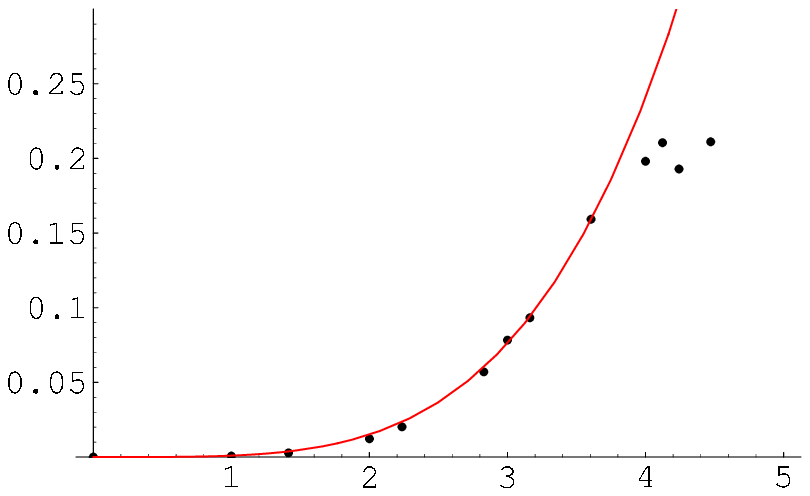}
\end{center}
\caption{In the left, the two lowest non-zero 
spectra at each momentum sector for the $L=5$ solution after one 
step of the renormalization are shown. 
The horizontal axis is $|p|=\sqrt{(p^1)^2+(p^2)^2}$.
In the right, the lowest spectra are fitted with $9.4 \times 10^{-4} |p|^4$. 
The departure of the spectra from the fitting line at $|p|\geq 4$ would 
be reasonable on account of the existence of the cutoff at $L=5$.
}
\label{fig:specL5ren}
\end{figure}

In Figure~\ref{fig:specL5ren}, the lowest four and two non-zero spectra at the 
vanishing and non-vanishing momentum sectors, resepctively,  
are plotted. 
The lowest trajectory can be well fitted with $|p|^4$ as previously. 
It is interesting to see that there seems to also exist
a massive trajectory.
The physical interpretation of the second trajectory is yet to be done. 

\section{Summary, conclusions and discussions}
In the previous paper \cite{Sasakura:2007sv}, the number distribution of the low-lying states around
the Gaussian solutions representing 
one- to four-dimensional fuzzy flat tori in a tensor model  was 
studied, and was shown to agree with the number distribution of 
the metric fluctuation modes transverse to the general coordinate transformation
in the general relativity. 
Although the agreement of the numbers for various dimensional tori is non-trivial,
it is certainly not enough for a definite conclusion. 
A disadvantage in the previous numerical study 
was the complexity of
the action, which made it difficult to go beyond the number counting. Therefore,  
in this paper, I have used a much simpler action having the Gaussian solutions, and 
have performed a detailed study of  
the  profiles of the low-lying modes around the Gaussian solutions representing two-dimensional fuzzy flat tori.
I have obtained very good matching with the metric fluctuation modes. 
The key issues in the comparison have been the proposal of the correspondence between 
the rank-three tensor in tensor models and the metric tensor in the general
relativity, and the DeWitt supermetric derived from the invariant metric
in tensor models under the correspondence.
It has also been observed that the low-lying modes are on a massless trajectory
with the dependence of $|p|^4$, 
which would be consistent with the fact in the general
relativity that the lowest momentum dependence will come from the $R^2$-term in 
two dimensions. The existence of such a well-shaped trajectory also implies 
the existence of a low-momentum effective smooth manifold.  
These results seem to lead to the definite conclusion that the low-momentum 
effective 
dynamics of the small fluctuations around the Gaussian solutions of tensor models is
described by the general relativity.

This conclusion insists that the symmetry $O(n)$ is enough as the
symmetry of tensor models to obtain the general relativity. This is 
rather unexpected, since the $GL(n,R)$ symmetry is more directly related to 
the general coordinate transformation \cite{Sasakura:2005js}. 
Probably, restricting only to the 
low-momentum dynamics, the distinction between the two symmetries may not be 
relevant. 
As future study,
this can be checked by investigating the properties of tensor models with dynamical
$g^{ab}$.

So far the study has been concentrated on the classical aspects, but 
more new interesting results are expected to come from 
the study of the quantum and thermodynamic properties, as in the Monte Carlo analysis of noncommutative field theories \cite{Panero:2006bx}.
A promising technical tool would be the renormalization procedure introduced in 
Section~\ref{sec:ren}. 
Even in the one-step classical application, the procedure has made the 
low-lying spectral patterns extensively simplified and clearer. 
This kind of simplifications will be expected to occur much more
in quantum mechanics and in thermodynamics, 
since quantum and thermodynamical fluctuations generally average over
detailed classical structures. 
Moreover, as shown in \eq{eq:renalpha}, the Gaussian solutions are right on the renormalization trajectories.
Therefore it can be expected that the renormalization procedure and 
the Gaussian solutions play much more important roles in quantum mechanics
and thermodynamics of tensor models than in the classical mechanics.
The optimistic hope is that the classification of tensor models finally reduces to 
that of actions having Gaussian solutions. 
Then the metric tensor in the general relativity can be regarded as the collective
dynamical variables applicable to  general tensor models  
under the correspondence \eq{eq:correspondence}.

In addition to the study of the dynamics, it would be interesting to study
the notion of nonassociative spaces \cite{Jackiw:1984rd}-\cite{Ho:2007vk} 
implicitly used behind, 
and also pursue philosophical basis in view of some known principles
and bounds in quantum gravity \cite{Garay}-\cite{Yoneya:2000bt}. It would also be interesting to analytically 
reproduce the results obtained so far by the brute-force numerical computations.  

\vspace{.5cm}
\section*{Acknowledgments}
The author was supported in part by the Grant-in-Aid for Scientific Research  No.16540244(C) and No.18340061(B)
from the Japan Society for the Promotion of Science (JSPS),
and also by the Grant-in-Aid for the 21st Century COE "Center for Diversity and Universality in Physics" from the Ministry of Education, Culture, Sports, Science and Technology (MEXT) of Japan.

\appendix

\section{Details of computation of spectra}
\label{sec:app}
Because of the momentum conservation of the background, the computation of 
spectra is easier in the momentum representation. From \eq{eq:defm},
taking into account $W_{abc}=0$ at a solution $C=C^0$, 
one obtains
\be
\label{eq:mww}
M^{(p_1,p_2,p_3),(p_1',p_2',p_3')}=\sum_{(p_4,p_5,p_6)}
\left.
\frac{m[(p_4,p_5,p_6)]}{\sqrt{m[(p_1,p_2,p_3)]m[(p_1',p_2',p_3')]}}
\frac{\partial W_{(p_4,p_5,p_6)}}{\partial C_{(p_1,p_2,p_3)}}
\frac{\partial W_{(-p_4,-p_5,-p_6)}}{\partial C_{(p_1',p_2',p_3')}}
\right|_{C=C^0}.
\ee

A slightly non-trivial matter in \eq{eq:mww} is the partial derivative of 
the fractional power of $K$ contained in $W$ with respect to $C$.
One has 
\be
\frac{\partial (K^{-\frac29})_a{}^b}{\partial C_{cde}}
=-(K^{-\frac29})_a{}^f \frac{\partial (K^{\frac29})_f{}^g}{\partial C_{cde}}
 (K^{-\frac29})_f{}^b.
\ee
In the momentum basis, the matrix $K^0{}_{a}{}^b$ ($K_a{}^b$ at
$C=C^0$) is diagonal on account of the momentum conservation. Moreover, from
the definition \eq{eq:defk}, the matrix is semi-positive definite and real symmetric
in the coordinate representation, 
and actually all the diagonal components are positive real at the 
numerical solutions
$C=C^0$. Therefore one can obtain a fractional
power of the matrix $K^0$ by taking the positive branch of the fractional powers of 
these positive diagonal components,
$X^0=(K^0)^\frac{l}{m}$, where $l,m$ are positive integers. 
Now let me consider its infinitesimal variation from $C=C^0$,
\be
(X^0+\delta X)^m=(K^0+\delta K)^l.
\ee
Taking the terms in the first order of $\delta X$ and $\delta K$ of the both sides,
and using the diagonal forms of $X^0$ and $K^0$, one obtains
\be
\label{eq:x0k0}
\delta X_a{}^b=\frac{X^0_a-X^0_b}{K^0_a-K^0_b} \delta K_a{}^b,
\ee
where $X^0_a$ and $K^0_a$ are the diagonal components of $X^0$ and $K^0$, 
respectively. If $K^0_a=K^0_b$, one can replace the fraction in \eq{eq:x0k0}
with its obvious limit, $\frac{l}{m} (K^0_a)^{\frac{l}{m}-1}$. Thus one obtains
\be
\label{eq:dkfdc}
\left.
\frac{\partial (K^\frac{l}{m})_a{}^b}{\partial C_{cde}}
\right|_{C=C^0}
=
\frac{(K^0_a)^\frac{l}{m}-(K^0_b)^\frac{l}{m}}{K^0_a-K^0_b} 
\left. \frac{\partial{K_a{}^b}}{\partial C_{cde}}\right|_{C=C^0}.
\ee

\begin{figure}
\begin{center}
\includegraphics[scale=.7]{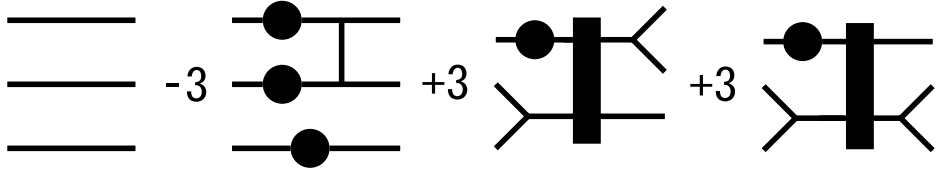}
\end{center}
\caption{The graphical expression of $\partial W_{abc}/\partial C_{def}$.
A three vertex denotes $C$, a blob denotes $K^{-\frac29}$, and 
a box denotes the fraction in \eq{eq:dkfdc}.
A line connecting these denotes a contraction of indices.
The multiplicity factors and the index symmetrization  
are abbreviated for simplicity.}
\label{fig:dwdc}
\end{figure}

The other partial derivatives with respect to $C$ are obvious.
The final expression of $\partial W/\partial C$ 
is graphically shown in Figure~\ref{fig:dwdc}, where
the multiplicity factors and the index symmetrization are abbreviated for 
simplicity\footnote{See the previous paper \cite{Sasakura:2007sv} for further details.}. 

\end{document}